\UseRawInputEncoding 
\documentclass[prd,aps, preprintnumbers,reprint,longbibliography,superscriptaddress,nofootinbib]{revtex4-1}
\usepackage{amsmath,amssymb,amsthm,amsxtra,overpic,bbm,bm,epsfig,subfigure}
\usepackage{hyperref}
\usepackage{mathrsfs}
\usepackage{enumitem}
\usepackage{graphicx}
\usepackage{comment}
\usepackage{epstopdf}
\usepackage{float}
\usepackage{multirow}
\usepackage{verbatim} 
\usepackage{color}
\usepackage{slashed}
\usepackage{multirow}
\usepackage{pstricks}
\usepackage{booktabs}
\usepackage{epstopdf}
\usepackage{bbding}
\usepackage{braket}
\usepackage{blkarray}

\newcommand{\prlsection}[2]{{\it\textbf{#1}{#2}}---}
\makeatletter
\newcommand*{\balancecolsandclearpage}{%
	\close@column@grid
	\cleardoublepage
	\twocolumngrid
}
\makeatother

\def \bal#1\eal  {\begin{align} #1 \end{align}}
\newcommand{\be} {\begin{equation}}
\newcommand{\ee} {\end{equation}}

\newcommand{\mc} {\mathcal}

\RequirePackage[normalem]{ulem} 
\RequirePackage{color}\definecolor{RED}{rgb}{1,0,0}\definecolor{BLUE}{rgb}{0,0,1} 

\begin{document}

\title{Origin of Neutrino Masses on the Convex Cone of Positivity Bounds}

\author{Xu Li}
\email{lixu96@ihep.ac.cn}

\author{Shun Zhou}
\email{zhoush@ihep.ac.cn (corresponding author)}

\affiliation{Institute of High Energy Physics, Chinese Academy of Sciences, Beijing 100049, China}
\affiliation{School of Physical Sciences, University of Chinese Academy of Sciences, Beijing 100049, China}

\date{\today}

\begin{abstract}
We exhibit the geometric structure of the convex cone in the linear space of the Wilson coefficients for the dimension-eight operators involving the left-handed lepton doublet $L$ and the Higgs doublet $H$ in the Standard Model effective field theory (SMEFT). The boundary of the convex cone gives rise to the positivity bounds on the Wilson coefficients, while the extremal ray corresponds to the unique particle state in the theory of ultra-violet completion. Among three types of canonical seesaw models for neutrino masses, we discover that only right-handed neutrinos in the type-I seesaw model show up as one of extremal rays, whereas the heavy particles in the type-II and type-III seesaw models live inside the cone. The experimental determination of the relevant Wilson coefficients close to the extremal ray of type-I seesaw model will unambiguously pin down or rule out the latter as the origin of neutrino masses. This discovery offers a novel way to distinguish the most popular seesaw model from others, and also strengthens the SMEFT as an especially powerful tool to probe new physics beyond the Standard Model.
\end{abstract}

\maketitle

\prlsection{Introduction}{.} The origin of neutrino masses is one of the tantalizing puzzles of Standard Model (SM), while it can be explained in the framework of the Standard Model effective field theory (SMEFT)~\cite{Buchmuller:1985jz, Grzadkowski:2010es, Brivio:2017vri} through the dimension-five (dim-5) Weinberg operator ${\cal O}^{(5)} \equiv \overline{L} \tilde{H} \tilde{H}^{\rm T} L^{\rm c}$~\cite{Weinberg:1979sa}. Therefore, it is very likely that new physics accounting for neutrino masses is connected to the left-handed lepton doublet $L$ and the Higgs doublet $H$. The type-I~\cite{Minkowski:1977sc, Yanagida:1979as, Gell-Mann:1979vob, Glashow:1979nm, Mohapatra:1979ia}, 
type-II~\cite{Konetschny:1977bn, Magg:1980ut, Schechter:1980gr, Cheng:1980qt, Mohapatra:1980yp, Lazarides:1980nt} 
and type-III~\cite{Foot:1988aq} seesaw models are tree-level ultra-violet (UV) completions of the Weinberg operator, but they are indistinguishable up to dim-5. To explore the origin of neutrino masses, one may turn to higher-dimensional operators in the corresponding low-energy effective field theories (EFT's)~\cite{Zhang:2021jdf,Ohlsson:2022hfl,Li:2022ipc,Du:2022vso}.

The positivity bounds are robust constraints on the Wilson coefficients (WC's) in the EFT, arising from the axiomatic principles of quantum field theories~\cite{Adams:2006sv,Pham:1985cr,Ananthanarayan:1994hf}. These bounds usually apply to dimension-eight (dim-8)~\cite{Bellazzini:2016xrt,Zhang:2018shp,Bellazzini:2018paj,Bi:2019phv,Remmen:2019cyz,Remmen:2020vts,Yamashita:2020gtt,Zhang:2020jyn, Fuks:2020ujk, Yamashita:2020gtt, Gu:2020ldn, Li:2021lpe,Du:2021byy,Grall:2021xxm,Chala:2021wpj, Zhang:2021eeo} or higher operators~\cite{deRham:2017avq,deRham:2017zjm,Tolley:2020gtv,Arkani-Hamed:2020blm,Bellazzini:2020cot,Caron-Huot:2020cmc,Chiang:2021ziz,Bellazzini:2021oaj,Alberte:2021dnj}, and have been used to narrow down the parameter space for experimental searches of new physics~\cite{Bellazzini:2018paj,Bi:2019phv,Yamashita:2020gtt,Fuks:2020ujk} or to test the fundamental principles of quantum field theories~\cite{Gu:2020ldn}. More attractively, the positivity for dim-8 operators is helpful in solving the ``inverse problem''~\cite{Arkani-Hamed:2005qjb, Dawson:2020oco, Gu:2020thj}, namely, reconstructing the UV theory from low-energy observables. This can be achieved from the geometrical perspective~\cite{Zhang:2020jyn, Fuks:2020ujk, Yamashita:2020gtt, Gu:2020ldn, Li:2021lpe,Du:2021byy, Zhang:2021eeo}, regarding the positivity bounds as normal vectors of a convex cone in the dim-8 WC space. The UV states living in the irreducible representations (irrep's) of  the symmetries of the $S$-matrix are projected into the WC space and form a convex hull. The heavy particles in the seesaw models are all in the irrep's of the SM gauge symmetry~\cite{Ma:1998dx}, implying a profound relation between the origin of neutrino masses and dim-8 physics. Solving the inverse problem provides us with a novel way to identify the seesaw model and understand the nature of neutrino masses. 

In this {\it letter}, we study the seesaw models of neutrino masses and establish for the first time the cone structure for the WC space of dim-8 operators containing $L$ and $H$. New positivity bounds on the $LLHH$ dim-8 operators are obtained, which are universal and accessible to future lepton colliders. We also solve the inverse problem to extract the UV information in the positivity region with the convex optimization algorithm.

\vspace{0.3cm}

\prlsection{Framework}{.} We focus on the second derivative of the forward 2-to-2 amplitudes ${\cal M}^{}_{ij \to kl}(s,t \to 0)$ with respect to $s$ (where $s,t $ are the ordinary Mandelstam variables, and the indices $i,j,k,l$ refer to the low-energy particles, including particle species, polarizations and quantum numbers), and introduce the tensor 
\begin{flalign}
	\begin{aligned}
		M^{ijkl} \equiv \lim_{s \to 0} \frac{d^2\mathcal{M}^{}_{ij\rightarrow k l}
			\left(s\right)}{ds^2} \; .
		\label{eq:tensor1}
	\end{aligned}
\end{flalign}
In the consideration of analyticity and unitarity, as well as the generalized optical theorem at the tree level, the dispersion relation in Eq.~(\ref{eq:tensor1}) can be recast into~\cite{deRham:2017avq, Bi:2019phv}
\begin{equation}
	\begin{aligned}
		M^{ijkl}&
		=\frac{1}{2\pi}\int_{(\varepsilon\Lambda)^2}^{\infty}{\frac{d\mu}{\mu^3}\sum_{X}\left[\mathcal{M}_{ij\rightarrow X}
			\mathcal{M}_{kl\rightarrow X}^\ast+\left(j\leftrightarrow l\right)\right]} \; ,
		\label{eq:tensor2}
	\end{aligned}
\end{equation}
where the summation is over all the intermediate UV states $X$ and the crossing channel is taken into account. Defining
$m^{ij} \equiv \mathcal{M}_{ij\rightarrow X}$, 
one can regard the result on the right-hand side of Eq.~(\ref{eq:tensor2}) as a positive linear combination of 
$m^{ij} m^{*kl}+(j\leftrightarrow l)$, since the integration can be understood as a limit of summation. Consequently, $M^{ijkl}$ belongs to the convex cone 
formed by $m^{ij} m^{*kl}+(j\leftrightarrow l)$, i.e.,
\begin{flalign}
	M^{ijkl}=\text{cone}\left\{m^{ij}m^{\ast k l}+m^{i\bar{l}}m^{\ast k\bar{j}}\right\}
	\label{eq:coneM}
\end{flalign}
where $\bar{j}$ and $\bar{l}$ denote the antiparticle states. For an EFT with $n$ low-energy states $i,j=1,...,n$, $m^{ij}$ by construction must be an $n$-dimensional matrix.

The extremal ray (ER) of the convex cone is defined as the element that cannot be decomposed into any nontrivial positive sum of other elements in the cone.  One observation from Eq.~(\ref{eq:tensor2}) is that the UV state residing in the irrep of the symmetry group corresponds to the ER of the cone~\cite{Zhang:2020jyn}. We choose $X$ to be the irrep $\mathbf{r}$ of the ${\rm SU}(3) \otimes {\rm SU}(2)^{}_{\rm L} \otimes {\rm U}(1)^{}_{\rm Y}$ gauge group, the particles $i$ and $j$ belong to the irrep $\mathbf{r}^{}_i$ and $\mathbf{r}^{}_j$, respectively. By the decomposition rule $\mathbf{r}^{}_i \otimes \mathbf{r}^{}_j = \sum_\alpha C_{\mathbf{r},\alpha}^{i,j} \mathbf{r}$, where $C_{\mathbf{r},\alpha}^{i,j}$ are the Clebsch-Gordan (CG) coefficients and the summation over all the states $\alpha$'s in $\mathbf{r}$ is implied, we can rewrite Eq.~(\ref{eq:tensor2}) as below
\begin{flalign}
	M^{ijkl}=\frac{1}{2\pi}\int_{\left(\varepsilon\Lambda\right)^2}^{\infty}{\frac{d\mu}{\mu^3}\sum_{X\text{\ in\ }\mathbf{r}}
		{\left|\left\langle X\middle|\mc{M}|\mathbf{r}\right\rangle\right|^2 \mc{G}_\mathbf{r}^{ijkl}}} \; ,
	\label{eq:projector}
\end{flalign}
with $\mc{G}_\mathbf{r}^{ijkl}\equiv \sum_\alpha C^{i,j}_{\mathbf{r},\alpha}
\left(C^{k,l}_{\mathbf{r},\alpha}\right)^\ast + (j \leftrightarrow l)$ 
being defined as the ``generator", which plays the role of an ER, and $M^{ijkl}$ can be generated by positive combinations of $\mc{G}_\mathbf{r}^{ijkl}$'s from different irrep's.
Extending the amplitudes $m^{ij}$ to include the CP-conjugate process, one can construct the generator as~\cite{Zhang:2021eeo}
\begin{equation}
	\begin{aligned}
		\mathcal{G}^{ijkl}=&m^{ij}m^{\ast k l}+m^{i\bar{l}}m^{\ast k\bar{j}}
		+m^{\bar{k}j}m^{\ast\bar{i}l} +m^{\bar{k}\bar{l}}m^{\ast\bar{i}\bar{j}}\\ &
		+\left(i\leftrightarrow 
		j,k\leftrightarrow l\right) \; ,  
		\label{eq:generator}
	\end{aligned}
\end{equation}
and thus the cone is defined by $\mathcal{C} = \text{cone}\left\{\mathcal{G}^{ijkl}\right\}$.

Now the primary goal is to examine the positions of seesaw models in the convex cone, and to identify them with the help of dim-8 operators. We notice the UV states of heavy particles in three types of seesaw models are in the irrep's of $\mathbf{1},\mathbf{3},\mathbf{3}$ of ${\rm SU}(2)^{}_{\rm L}$ gauge symmetry, respectively, so they naturally fit into this framework. Given all known symmetries in the theory, it is straightforward to find all UV states that lead to dim-8 operators.

\vspace{0.3cm}

\prlsection{The UV States}{.} As mentioned before, nonzero neutrino masses may hint at the existence of the Weinberg operator  ${\cal O}^{(5)} \equiv \overline{L} \tilde{H} \tilde{H}^{\rm T} L^{\rm c}$ and thus new physics associated with lepton and Higgs doublets. To gain more information about possible UV states, we examine the minimal space of WC's, including all four-particle operators ${\cal O}^{}_{ijkl}$ with $i,j,k,l=H$ or $L$. For simplicity, only one lepton flavor is considered, but the extension to three flavors is straightforward. The dim-8 operators that contribute to the forward scattering amplitudes with the $s^2$-dependence can be classified into three types of subspaces~\cite{Li:2020gnx, Murphy:2020rsh}:
\begin{itemize}
	\item  $LLHH$: 
	\begin{equation}
		\begin{aligned}
			&{\cal O}^{}_1= (\bar{L}\gamma_\mu {\rm i} \overleftrightarrow{D_\nu} L ) \left(D^\mu H^\dag  D^\nu H\right), \\
			&{\cal O}^{}_2= (\bar{L}\gamma_\mu\sigma^I {\rm i} \overleftrightarrow{D_\nu} L) \left(D^\mu H^\dag\sigma^I D^\nu H\right) \; ;
			\label{eq:LLHH}
		\end{aligned}
	\end{equation}
	\item  $LLLL$: 
	\begin{equation}
		\begin{aligned}
			&{\cal O}^{}_3 = \partial_\nu\left(\bar{L}\gamma^\mu L\right)\partial^\nu\left(\bar{L}\gamma_\mu L\right),\\
			&{\cal O}^{}_4 = \partial_\nu\left(\bar{L}\gamma^\mu\sigma^I L\right)\partial^\nu\left(\bar{L}\gamma_\mu\sigma^I L\right) \; ;
		\end{aligned}
	\end{equation}
	\item $HHHH$: 
	\begin{equation}
		\begin{aligned}
			&{\cal O}^{}_5=\left(D_\mu H^\dag D_\nu H\right)\left(D^\nu H^\dag D^\mu H\right),\\
			&{\cal O}^{}_6=\left(D_\mu H^\dag D_\nu H\right)\left(D^\mu H^\dag D^\nu H\right),\\
			&{\cal O}^{}_7=\left(D_\mu H^\dag D^\mu H\right)\left(D_\nu H^\dag D^\nu H\right) \; .
		\end{aligned}
	\end{equation}
\end{itemize}
where $\sigma^I$ (for $I = 1, 2, 3$) stand for the Pauli matrices and $\overleftrightarrow{D_\nu} \equiv D^{}_\mu - \overleftarrow{D^{}_\mu}$ with $D^{}_\mu$ being the covariant derivative in the SM has been defined.  
Note that there are another two dim-8 operators in the subspace of $LLHH$, namely, $(\bar{L}\gamma_\mu  L ) \square  ( H^\dag {\rm i} \overleftrightarrow{D^\mu} H )$ and $(\bar{L}\gamma_\mu \sigma^I L ) \square ( H^\dag \sigma^I  {\rm i} \overleftrightarrow{D^\mu}  H )$. For the forward scattering $H(p_1) L(p_2)\to H(p_1) L(p_2)$, where the four-momenta are specified, the d'Alembert operators will produce $p_1^2=0$, leading to a vanishing amplitude. On the other hand, for the $k \leftrightarrow l$ exchanged process $H(p_1) L(p_2)\to L(p_1) H(p_2)$, the covariant derivative $D^\mu$ acting on the Higgs field $H$ provides a $p_1^\mu$ or $p_2^\mu$ factor, which will be contracted with $\gamma^{}_\mu$ in the fermion current and renders the amplitude to vanish. Due to the crossing symmetry, other amplitudes induced by these two dim-8 operators also vanish. Therefore, they don't contribute to the amplitudes of our interest. 

The positivity bounds in two subspaces of $HHHH$ and $LLLL$ have been studied in the literature~\cite{Remmen:2019cyz, Zhang:2020jyn, Fuks:2020ujk, Zhang:2021eeo}. In the present {\it letter}, we enlarge the space of WC's by further combining those two operators ${\cal O}^{}_1$ and ${\cal O}^{}_2$ in Eq.~(\ref{eq:LLHH}). As we shall show later, the results obtained in the previous works can be reproduced when restricted to the subspace $HHHH$ or $LLLL$.
\begin{table*}
	\renewcommand\arraystretch{1.2}
	\setlength{\tabcolsep}{1.2mm}
	\begin{tabular}{ c c c c c c c }
		\hline \hline UV State & Spin& ${\rm SU}(2)^{}_{\rm L} \otimes {\rm U}(1)^{}_{\rm Y}$ & Interaction & Seesaw &  Extremal Ray & $\vec{c}$ \\
		\hline \hline$E$ & $1/2$ & ${\bf 1}_{-1}$ & $g \bar{E}\left(H^{\dagger} L\right)$ & & \CheckmarkBold &$\displaystyle \frac{1}{2}(-1,-1,0,0,0,0,0)$\\
		$\Sigma_1$ & $1/2$&$ {\bf 3}_{-1}$ & $g \bar{\Sigma}^I_1 \left(H^{\dagger} \sigma^{I} L\right)$ 
		& & \XSolidBrush &$\displaystyle \frac{1}{2}(-3,1,0,0,0,0,0)$\\
		$N$ & $1/2$&$ {\bf 1}_{0}$ & $g \bar{N}\left(H^{\rm T} \epsilon L\right)$ &$\operatorname{Type-I}$ & \CheckmarkBold
		&$\displaystyle \frac{1}{2}(-1,1,0,0,0,0,0)$ \\
		$\Sigma$ & $1/2$&$ {\bf 3}_{0}$ & $g \bar{\Sigma}^I \left(H^{\rm T} \epsilon \sigma^{I} L\right)$ 
		& $\operatorname{Type-III}$ & \XSolidBrush & $\displaystyle \frac{1}{2}\left(-3,-1,0,0,0,0,0\right)$\\
		$\mathcal{B}_1$ & $1$&${\bf 1}_{1}$ & $g \mathcal{B}_1^{\mu} \left[(H^{\dag} \epsilon {\rm i} \overleftrightarrow{D_{\mu}} H^\ast)
		+ \frac{x}{M} (\bar{L^{\rm c}} \epsilon {\rm i} \overleftrightarrow{D_{\mu}} L)\right]$ 
		& &  \XSolidBrush &$\displaystyle \frac{1}{2}(0,0,x^2,-x^2,16,0,-16)$\\
		$\Xi_1$ & $0$&${\bf 3}_{1}$ & $g \Xi^{I}_1 \left[M (H^{\dag} \epsilon \sigma^{I} H^\ast )+ x (\bar{L^{\rm c}} 
		\epsilon \sigma^{I} L )\right]$ & $\operatorname{Type-II}$& \XSolidBrush & $\displaystyle \frac{1}{2}\left(0,0,-3 x^{2}, - x^{2},0,16,0\right)$ \\
		$\mathcal{S}$ & $0$ & ${\bf 1}_{0 \rm S}$ & $g M \mathcal{S}\left(H^{\dagger} H\right)$ & & \CheckmarkBold  &$2(0,0,0,0,0,0,1)$\\
		$\mathcal{B}$ & $1$ & ${\bf 1}_{0 \rm A}$ & $g \mathcal{B}^{\mu}\left[H^{\dagger}{{\rm i}\overleftrightarrow{D_{\nu}}} 
		H+x (\bar{L} \gamma_{\mu} L)\right]$ & &\XSolidBrush  &$\displaystyle \frac{1}{2}\left(0,0,-x^{2},0, -4,4,0\right)$ \\
		$\Xi$ & $0$ & ${\bf 3}_{0 \rm S}$ & $g M \Xi^{I} (H^{\dagger} \sigma^{I} H )$ & & \XSolidBrush & $2(0,0,0,0,2,0,-1)$\\
		$\mathcal{W}$ & $1$ & ${\bf 3}_{0 \rm A}$ & $g \mathcal{W}^{I \mu}\left[ (H^{\dagger} \sigma^{I} 
		{\rm i}\overleftrightarrow{D_{\mu}} H )
		+ x (\bar{L} \gamma_{\mu} \sigma^{I} L)\right]$ & &\XSolidBrush & $\displaystyle \frac{1}{2}\left(0,0,0,-x^{2},4,4,-8\right)$\\
		\hline\hline
	\end{tabular}
	\vspace{0.5cm}
	\caption{All possible UV states that can be coupled to the lepton doublet $L$ and the Higgs doublet $H$, where the spin, the representations of the ${\rm SU}(2)^{}_{\rm L} \otimes {\rm U}(1)^{}_{\rm Y}$ group [the subscripts ``S" and ``A" refer to symmetric and antisymmetric representations in Eq.~(\ref{eq:mmatrix})] and the corresponding vectors in the $LLHH$
		subspace of Wilson coefficients are shown. Notice that $g$ is the coupling constant, $x$ is an arbitrary real number characterizing the relative size of couplings, and $M$ is the mass parameter.}
	\label{tab:UVstates}
\end{table*}

Since both $L$ and $H$ are assigned as the irrep $\mathbf{2}$ of the ${\rm SU}(2)^{}_{\rm L}$ group in the complex-field basis, the CG coefficients of direct product decomposition
$ \mathbf{2}\times \mathbf{2}= \mathbf{1}+\mathbf{3}$ and 
$ \mathbf{2}\times \bar{\mathbf{2}}= \mathbf{1}+\mathbf{3}$, labeled 
by $C_{\mathbf{1}/\mathbf{3},c}^{ab}$ and ${\bar{C}}_{\mathbf{1}/\mathbf{3},c}^{ab}$, read
\begin{equation}
	\begin{aligned}
		&C_{\mathbf{1},c}^{ab} = \epsilon^{ab} \; , \quad C_{\mathbf{3},c}^{ab} = \left(\epsilon \sigma^I\right)^{ab} \; ; \\
		&{\bar{C}}_{\mathbf{1},c}^{ab} = \delta_b^a \; , \quad \, {\bar{C}}_{\mathbf{3},c}^{ab} = \left(\sigma^I\right)^a_b \; ,
	\end{aligned}
\end{equation}
where $\epsilon \equiv {\rm i}\sigma^2$ and the subscript ``$c$" is trivial for ${\bf 1}$ but $c = I$ for ${\bf 3}$. With those CG coefficients, the $m^{ij}$ matrix can be  
found immediately, i.e.
\begin{flalign}
	\raisebox{-10pt}{$m= $}&
	\renewcommand\arraystretch{1.5} 
	\setlength{\tabcolsep}{1.5mm}
	\begin{tabular}{ r|c|c|c|c| }
		\multicolumn{1}{r}{}
		&  \multicolumn{1}{c}{$H_b$}
		& \multicolumn{1}{c}{$L_b$}
		&  \multicolumn{1}{c}{$ H^\dagger_b $}
		& \multicolumn{1}{c}{$ \bar{L}_b $}  \\
		\cline{2-5}
		$ H_a $&  $C_{\mathbf{1}/\mathbf{3},c}^{ab}$ &  $C_{\mathbf{1}/\mathbf{3},c}^{ab}$  & ${\bar{C}}_{\mathbf{1}/\mathbf{3},c}^{ab}$ 
		& ${\bar{C}}_{\mathbf{1}/\mathbf{3},c}^{ab}$  \\
		\cline{2-5}
		$ L_a $& $C_{\mathbf{1}/\mathbf{3},c}^{ab}$  & $xC_{\mathbf{1}/\mathbf{3},c}^{ab}$  &  $C_{\mathbf{1}/\mathbf{3},c}^{ab}$ & $x{\bar{C}}_{\mathbf{1}/\mathbf{3},c}^{ab}$   \\
		\cline{2-5}
		$ H^\dagger_a $ &  $\pm{\bar{C}}_{\mathbf{1}/\mathbf{3},c}^{ab}$  & $C_{\mathbf{1}/\mathbf{3},c}^{ab}$ & $C_{\mathbf{1}/\mathbf{3},c}^{ab}$ & $C_{\mathbf{1}/\mathbf{3},c}^{ab}$ \\
		\cline{2-5}
		$ \bar{L}_a $& ${\bar{C}}_{\mathbf{1}/\mathbf{3},c}^{ab}$  & ${\rm i} x{\bar{C}}_{\mathbf{1}/\mathbf{3},c}^{ab}$  & $C_{\mathbf{1}/\mathbf{3},c}^{ab}$  & $x C_{\mathbf{1}/\mathbf{3},c}^{ab}$  \\
		\cline{2-5}
	\end{tabular} \; ,
	\label{eq:mmatrix}
\end{flalign}
where $x$ is an arbitrary real parameter, representing the relative size of the coupling constant between $X$ and $HH$ (or $H^\dag H$) to that between $X$ and $LL$ (or $\bar{L}L$). The generator  $\mc{G}^{ijkl}_{\mathbf{r}}$ can be derived from Eq.~(\ref{eq:generator}) for each irrep $\mathbf{r}$, and will be matched into the WC space by identifying $\mathcal{G}^{ijkl}_{\mathbf{r}} = M^{ijkl}\left(C_1, C_2,\cdots, C_6, C_7\right)$. In fact, it can be effectively viewed as a vector $\vec{c}_{\mathbf{r}}$ in the WC space. On the other hand, all the generators can be interpreted as the tree-level exchange of a single heavy state $X$ in the irrep $\mathbf{r}$. Therefore, one can evaluate $M^{ijkl}$ in the UV theory and integrate $X$ out to match $M^{ijkl}$ into the WC space. In this way, another vector will be obtained, but it must be identical to $\vec{c}_{\mathbf{r}}$ up to an overall positive factor. 

In Table~\ref{tab:UVstates}, we list all possible scenarios of tree-level UV completion. In each scenario, the single heavy state is specified with the spin, the quantum number ${\bf r}^{}_{Y}$ under the ${\rm SU}(2)^{}_{\rm L}\otimes {\rm U}(1)^{}_{\rm Y}$ group, the interactions with light SM particles, and the corresponding vector $\vec{c}$ in the WC space. Once all those vectors in the WC space are obtained, the ER's will be identified the subset of $\vec{c}$'s so that other vectors can be positively decomposed into the ER's. We also explicitly indicate which vector is ER in one column. In particular, three types of seesaw models for neutrino masses are also marked.

Some comments on Table~\ref{tab:UVstates} are in order. First, the subspace of $LLHH$ formed by the WC's $C_1$ and $C_2$ is orthogonal to two other types of subspaces, i.e., $LLLL$ and $HHHH$. In the $LLHH$ subspace, there are four scenarios of UV completion, for which all the UV states are fermions. In other scenarios, the UV states are bosons. This is reasonable since one fermion cannot mediate interactions between two $L$'s or two $H$'s in the $s$-channel. Meanwhile, for the forward scattering, one boson cannot be tree-level UV completion of the effective vertices in ${\cal O}^{}_1$ and ${\cal O}^{}_2$. Otherwise, the resultant amplitude takes the form of $p_2^2 \bar{v}(p^{}_1) \slashed{p}^{}_1 u(p^{}_2)$ with $v(p^{}_1)$ and $u(p^{}_2)$ being the wave functions of external leptons (with four-momenta $p^{}_1$ and $p^{}_2$), which would vanish after applying equations of motion. This observation allows us to discuss the $LLHH$ subspace without worrying about the other WC's in the next section. 

Second, if we restrict ourselves into the $HHHH$ subspace, corresponding to the last three components of $\vec{c}$, the results in Refs.~\cite{Remmen:2019cyz, Zhang:2020jyn} can be reproduced with the positivity bounds $C_6\geq 0,C_5+C_6\geq 0,C_5+C_6+C_7\geq 0$. Similarly, when reduced to the $LLLL$ subspace, the same conclusions of the positivity bounds $C_3+C_4\leq 0,C_4\leq 0$ in Refs.~\cite{Fuks:2020ujk, Zhang:2021eeo} are reached . 

Finally, it is worthwhile to notice that the $\vec{c}$'s in the subspaces $LLLL$ and $HHHH$ don't contain the components linear in $x$. Since the generator $\mc{G}$ is the product of two $m$'s, while $m^{ij}$ is the linear function of $x$, it is expected that $\vec{c}$'s have the components both linearly and quadratically dependent on $x$, and independent of $x$. The reason is simply that the components linear in $x$ can only appear in the $LLHH$ subspace, but the one-boson realization of the operators ${\cal O}^{}_1$ and ${\cal O}^{}_2$ is impossible. This feature guarantees that the ER's in the subspace $LLLL$ and $HHHH$, respectively, remain to be so in the enlarged space, keeping the previously derived positivity bounds intact.

\vspace{0.3cm}
\begin{figure}[htb]
	\begin{center}
		\includegraphics[width=0.7\linewidth]{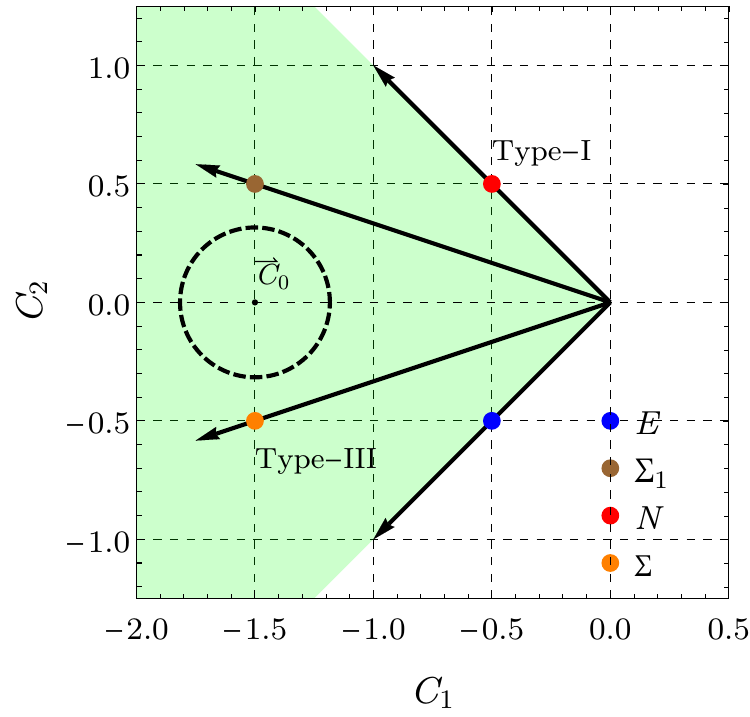}
	\end{center}
	\caption{The geometric structure of the convex cone in the $LLHH$ subspace, where four different scenarios of UV completion are represented by the vectors with four colored dots. The length of the vector in the $(C_1,C_2)$-plane represents the amplitude from the tree-level matching, with the coupling in the interaction Lagrangian in Table \ref{tab:UVstates} set to $g = 1$ and a common mass for the UV state. Only the green region is allowed by positivity bounds, and the vector $\vec{C}_0 = (-3/2, 0)$ is a benchmark point of experimental measurement with the dashed circle being the associated uncertainties.
	}
	\label{fig:2Dspace}
\end{figure}

\begin{figure}[htb]
	\begin{center}
		\includegraphics[width=\linewidth]{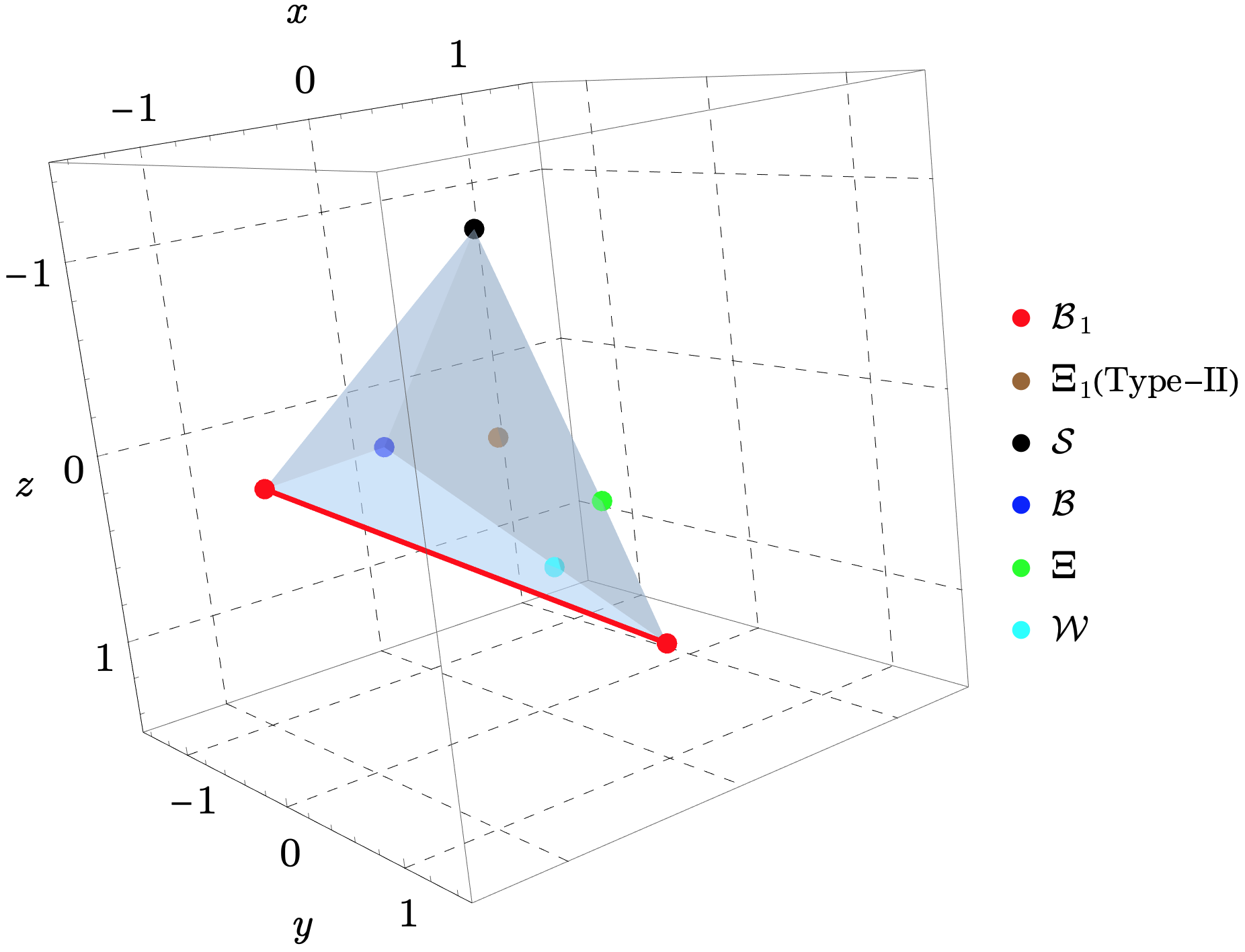}
	\end{center}
	\caption{The 3D cross section of the 5D $LLLL+HHHH$ subspace. To best visualize the $\mc{B}_1^{}$ state represented by the red line, we have chosen a particular direction to project the UV states with $(x,y,z)=(-0.7 C_3+0.69 C_4+0.11 C_5+0.13 C_6+0.057 C_7,0.031 C_3+0.85 C_5-0.5 C_6-0.16 C_7, -0.019 C_3+0.31 C_6-0.95 C_7)/(C_3+2C_4-2C_5-3C_6-C_7)$, and the UV states $(\Xi_1, \mc{B},\mc{W}^I)$ are projected as points by specifying $x^2=(1/4,1,1/4)$.
	}
	\label{fig:3Dspace}
\end{figure}

\prlsection{The LLHH Subspace}{.} In the two-dimensional $LLHH$ subspace, the $x$- and $y$-axis actually refer to $C^{}_1$ and $C^{}_2$, respectively. This subspace is orthogonal to the rest and thus we simply set $C_3, \cdots , C_7 = 0$ for the moment. The first four rows in Table~\ref{tab:UVstates} correspond to the UV completion with fermions, and the corresponding vectors $\vec{c}$'s are shown in Fig.~\ref{fig:2Dspace}. The polyhedron region in green is allowed by positivity bounds. The ER's are the two edges of polyhedron, namely,  $\vec{c}=(-1/2, -1/2)$ and $(-1/2,1/2)$. The positivity bounds are just the normal vectors of the two edges, i.e.
\be
C_1 + C_2  \leq 0,\quad  C_1 - C_2  \leq 0 \; .
\ee
which are first derived here. Interestingly, we find that the type-I and type-III seesaw models belong to this subspace, but only the type-I seesaw
lives on one of two edges. On the other hand, the type-II seesaw lives in the five-dimensional subspace of $LLLL$ and $HHHH$ and appears in the convex cone, as shown in its three-dimensional projection in Fig.~\ref{fig:3Dspace}. To better visualize the convex cone and the UV models, we have chosen a particular two-dimensional direction as explained in the caption of Fig.~\ref{fig:3Dspace}.
Another practical application of the convex geometry is to solve the inverse problem~\cite{Fuks:2020ujk, Zhang:2021eeo}, as any UV completion must have net dim-8 effects that cannot be completely lifted by the contributions from other possible UV completions. Then we proceed to explain how to infer the information about the UV physics. Once the collider experiments observe the benchmark point that fixes the vector $\vec{C} = (-3/2, 0)$, it should be a positive combination of the generator vectors, i.e., $\vec{C} = \sum_i \omega_i \vec{c}_i$ with $i=E, N, \Sigma, \Sigma_1$ and $c_i$ being the vector corresponding to each UV state in the $C_1$-$C_2$ plane. The coefficients $\omega_i = g_i^2/M_i^4$ are positive, and they carry the very information about the UV theory, namely, the relevant couplings and masses.

For instance, if the measured data point is located exactly on the edge represented by $\vec{c}=(-1/2,1/2)$, then one can pin down the existence of $N$, i.e., the UV state in the type-I seesaw. At the same time, the existence of other UV states $E,\Sigma,\Sigma_1$ can be excluded. These conclusions are guaranteed by the salient feature of the ER of the convex cone. If the data point lies on the edge, the associated vector cannot be decomposed into any other vectors. Therefore, the only possible UV state $X$ should be the one in the irrep $\mathbf{r}$ corresponding to that edge.

Generally, the measured data point may be not on the edge but inside the cone. In reality, the experimental results of the WC's are usually reported as a region bounded by the multidimensional ellipsoid, which is determined by the $\Delta \chi^2$-value. Then the question is how to extract the constraints on $\omega_i$  from experimental data. The solution has been provided in Refs.~\cite{Fuks:2020ujk,Zhang:2021eeo}.
If the experimental result is represented by a point $\vec{C}_0$ in the $C_1$-$C_2$ plane, then the upper bound on $\omega_i$ can be derived by finding the maximal value $\lambda^{}_{\rm max}$ of $\lambda$ such that the following vector breaks the positivity condition
\be
\vec{C}(\lambda) \equiv \vec{C}_0-\lambda \vec{c_i} = \sum_{j \neq i} \omega_j \vec{c}_j + (\omega_i-\lambda) \vec{c}_i \; .
\ee
The value of $\lambda$ can be stated as the maximum possibility for the UV state $i$ to exist and explain the experimental data. Quantitatively, the upper bound on $\omega_i$ is given by $\lambda_{\rm max}$, i.e., $\lambda_{\rm max}\geq g_i^2/M_i^4$. Unlike the numerical solution in Refs.~\cite{Fuks:2020ujk, Zhang:2021eeo}, we find that this can be identified as a conic optimization problem, thanks to the convex nature of the WC space. 

Given the uncertainty as a multidimensional ellipsoid, the upper bound on $\omega_i$ can be determined since the conic optimization reduces to the second-order cone program
\be
\begin{aligned}
	\text{maximize}& \quad \lambda \\
	\text{subject to}& \quad \vec{C}-\lambda \vec{c_i} \subset \mc{C}\\
	&\quad (\vec{C}-\vec{C_0}) \cdot A \cdot (\vec{C}-\vec{C_0}) \leq \Delta
	\label{eq:ConicOpti}
\end{aligned}
\ee 
where $A$ is the covariant matrix from the $\chi^2$-analysis, $\vec{C}_0$ is the best-fit point, and $\Delta$ is determined by the desired confidence level and by the number of free parameters. If the $\Delta$ constraints are absent, the problem automatically reduces to the linear optimization program. Both these two optimization problems can be solved by the well-established computer algorithms.
\begin{table}
	\renewcommand\arraystretch{1.2}
	\setlength{\tabcolsep}{1.2mm}
	\begin{tabular}{ c c c c c }
		\hline \hline $\vec{C}_0$ & $E$ & $\Sigma_1$ & $N$ & $\Sigma$  \\
		\hline \hline
		$(-1/2,1/2)$ & $\infty$  & $\infty$ & $ \geq 1.0 $ & $\infty$ \\
		$(-3/2,0)$ & $ \geq 0.9 $ & $\geq 1.07 $ & $\geq 0.9 $ & $\geq 1.07 $\\
		$(-3/2,0)~\text{with}~\Delta = 0.1$ & $\geq 0.85 $ & $\geq 1.0 $ & $\geq 0.85 $ &$\geq 1.0 $ \\
		$(0,0)~\text{with}~\Delta = 0.1$  & $\geq 1.22 $ & $\geq 1.5 $ & $\geq 1.22 $ & $\geq 1.5 $ \\
		\hline\hline
	\end{tabular}
	\vspace{0.5cm}
	\caption{The derived lower bounds on $M^{}_i/\sqrt{g_i}$ in units of TeV for each UV state with $i=E,N,\Sigma,\Sigma_1$. The measured data are represented by two points in the first two rows, whereas by the allowed ranges in Eq.~(\ref{eq:ConicOpti}) with $\Delta = 0.1$ in the last two rows.}
	\label{tab:bounds}
\end{table}

For illustration, we take the best-fit point $\vec{C}_0 = (-3/2,0)$ and the constraint as the disc  $(C_1+3/2)^2 + C_2^2 \leq 0.1$, whose boundary has been plotted as the dashed circle in Fig.~\ref{fig:2Dspace}. In Table~\ref{tab:bounds}, we summarize the results by solving Eq.~(\ref{eq:ConicOpti}) in such a simple setup. The bounds on $\omega^{}_i$ for the benchmark point  $\vec{C}_0 = (-3/2, 0)$ and that for the point on the edge $\vec{C}_0 = (-1/2,1/2)$ have been derived and then converted into the bounds on $M^{}_i/\sqrt{g_i}$ in units of TeV for each UV state. In the former case, it is difficult to solve the inverse problem, i.e., all the UV models fit the measurement equally well. But, in the latter case, the type-I seesaw model is singled out even if the experimental uncertainty is taken into account. In contrast, if the experimental results point to $\vec{C_0} = (0, 0)$, all the UV models will unambiguously be ruled out up to a certain mass scale.

\vspace{0.3cm}

\prlsection{Summary}{.} Motivated by nonzero neutrino masses observed in neutrino oscillation experiments, we stress that the Weinberg operator ${\cal O}^{(5)} \equiv \overline{L} \tilde{H} \tilde{H}^{\rm T} L^{\rm c}$ for tiny Majorana neutrino masses may naturally exist in the SMEFT and new physics beyond the SM is very likely connected to the lepton and Higgs doublets. Therefore, we examine three classes of dim-8 operators involving lepton and Higgs doublets and reveal the geometric structure of the convex cone of positivity bounds in the subspace of the relevant WC's at the tree level. The discussions about positivity bounds at the one-loop level can be found in Ref.~\cite{Chala:2021wpj}.

In the subspace of the WC's for two $LLHH$ operators, we discover that the type-I seesaw model resides on the edge of the convex cone, indicating that the measurement of the WC's close to the edge will unambiguously confirm or rule out the type-I seesaw model as the true theory of neutrino masses. However, type-II and type-III seesaw models live inside the convex cone. This discovery provides a new and highly nontrivial way to distinguish between the type-I seesaw model and its analogues. We also explain how to extract the constraints on the UV theories once the experimental measurements of the WC's of dim-8 operators are available.  

Obviously the key point is to experimentally measure the relevant WC's of dim-8 operators in the $LLHH$ class. More and more data will be accumulated at the CERN Large Hadron Collider and future lepton or hadron colliders, offering the possibility to probe dim-8 operators~\cite{Li:2020gnx, Murphy:2020rsh, Liu:2016idz, Azatov:2016sqh, Ellis:2018cos, Hays:2018zze, Bellazzini:2018paj, Ellis:2019zex, Ellis:2020ljj, Alioli:2020kez, Remmen:2020vts}. For example, one can probe the pair production of the Higgs bosons via $e^+ e^- \to h h$ in future electron-positron colliders~\cite{Vasquez:2019muw}, 
to determine the values of $C_1$ and $C_2$.
As tree-level SM contributions will be highly suppressed by the electron Yukawa coupling and the Higgs self-coupling, this channel may be an ideal place to test the effects of dim-8 operators. Based on the studies in Ref.~\cite{Vasquez:2019muw}, observing the Higgs pair production at a lepton collider requires the center-of-mass energy $\sqrt{s}$ to be higher than $400$ GeV, which can be reached at the linear colliders, e.g., the Compact Linear Collider (CLIC)\cite{CLIC:2018fvx}. Given the nominal setup $\sqrt{s}=1.5$ TeV and the integrated luminosity ${\cal L} =1.5\ {\rm ab}^{-1}$ at the CLIC, one can roughly estimate the sensitivity as $|C_{\text{dim-8}}/\Lambda^4|\leq \mc{O}(10^{-2})\ {\rm TeV}^{-4}$ at the $95\%$ confidence level. Further detailed studies in this direction are interesting and desirable.

\vspace{0.3cm}

\prlsection{Acknowledgements}{.} One of the authors (X.L.) would like to thank the late Prof. Cen Zhang for early helpful discussions, and both authors are greatly indebted to Prof. Zhi-zhong Xing and Prof. Shuang-Yong Zhou for their valuable suggestions. This work was supported in part by the National Natural Science Foundation of China under grant No.~11835013 and the Key Research Program of the Chinese Academy of Sciences under grant No. XDPB15.

\bibliography{refs}
\bibliographystyle{JHEP}

\end{document}